\documentclass[%
 reprint,
 superscriptaddress,
 amsmath,amssymb,
 aps,
prl,
]{revtex4-2}

\usepackage{graphicx}% Include figure files
\usepackage{dcolumn}% Align table columns on decimal point
\usepackage{bm}% bold math
\usepackage[mathlines]{lineno}% Enable numbering of text and display math
\begin{document}

\preprint{APS/123-QED}

\title{Revealing the nature of yrast states in neutron-rich polonium isotopes}

\author{R.~Lic\u{a}}
 \email{razvan.lica@nipne.ro}
 \affiliation{Horia Hulubei National Institute for R\&D in Physics and Nuclear Engineering, RO-077125 Bucharest, Romania}
 \author{A.N.~Andreyev}
 \affiliation{School of Physics, Engineering and Technology, University of York, York, YO10 5DD, United Kingdom}
\author{H.~Na\"idja}
 \affiliation{Universit\'e Constantine 1, Laboratoire de Physique Math\'ematique et Physique Subatomique, Constantine 25000, Algeria}
\author{A.~Blazhev}
 \affiliation{Institut f{\"u}r Kernphysik, Universit{\"a}t zu K{\"o}ln, D-50937 K{\"o}ln, Germany}
\author{P.~Van~Duppen}
\affiliation{KU Leuven, Instituut voor Kern- en Stralingsfysica, B-3001 Leuven, Belgium}
\author{B.~Andel}
\affiliation{Department of Nuclear Physics and Biophysics, Comenius University in Bratislava, 84248 Bratislava, Slovakia}

\author{A.~Algora}
\affiliation{Instituto de F{\'i}sica Corpuscular, CSIC - Universidad de Valencia, E-46980, Valencia, Spain}
\affiliation{Institute of Nuclear Research (ATOMKI), P.O.Box 51, H-4001 Debrecen, Hungary}
\author{S.~Antalic}
\affiliation{Department of Nuclear Physics and Biophysics, Comenius University in Bratislava, 84248 Bratislava, Slovakia}
\author{J.~Benito}
\affiliation{Grupo de F\`isica Nuclear, EMFTEL \& IPARCOS, 
Universidad Complutense de Madrid, 28040 Madrid, SPAIN }
\author{G.~Benzoni}
\affiliation{Istituto Nazionale di Fisica Nucleare, Sezione di Milano, I-20133 Milano, Italy}
\author{T.~Berry}
\affiliation{Department of Physics, University of Surrey, Guildford GU2 7XH, United Kingdom}
\author{M.~J.~G.~Borge}
\affiliation{Instituto de Estructura de la Materia, CSIC, Serrano 113 bis, E-28006 Madrid, Spain}
\author{C.~Costache}
\affiliation{Horia Hulubei National Institute for R\&D in Physics and Nuclear Engineering, RO-077125 Bucharest, Romania}
\author{J.~G.~Cubiss}
\affiliation{School of Physics, Engineering and Technology, University of York, York, YO10 5DD, United Kingdom}
\author{H.~De~Witte}
\affiliation{KU Leuven, Instituut voor Kern- en Stralingsfysica, B-3001 Leuven, Belgium}
\author{L.~M.~Fraile}
\affiliation{Grupo de F\`isica Nuclear, EMFTEL \& IPARCOS, 
Universidad Complutense de Madrid, 28040 Madrid, SPAIN }
\author{H.~O.~U.~Fynbo}
\affiliation{Department of Physics and Astronomy, Aarhus University, DK-8000 Aarhus C, Denmark}
\author{P.~T.~Greenlees}
\affiliation{University of Jyv\"askyl\"a, Department of Physics, Accelerator laboratory, P.O. Box 35(YFL) FI-40014 University of Jyv\"askyl\"a, Finland}
\author{L.~J.~Harkness-Brennan}
\affiliation{Department of Physics, Oliver Lodge Laboratory, University of Liverpool, Liverpool L69 7ZE, United Kingdom}
\author{M.~Huyse}
\affiliation{KU Leuven, Instituut voor Kern- en Stralingsfysica, B-3001 Leuven, Belgium}
\author{A.~Illana}
\affiliation{Instituto Nazionale di Fisica Nucleare, Laboratori Nazionali di Legnaro, I-35020 Legnaro, Italy}
\author{J.~Jolie}
\affiliation{Institut f{\"u}r Kernphysik, Universit{\"a}t zu K{\"o}ln, D-50937 K{\"o}ln, Germany}
\author{D.~S.~Judson}
\affiliation{Department of Physics, Oliver Lodge Laboratory, University of Liverpool, Liverpool L69 7ZE, United Kingdom}
\author{J.~Konki}
\affiliation{University of Jyv\"askyl\"a, Department of Physics, Accelerator laboratory, P.O. Box 35(YFL) FI-40014 University of Jyv\"askyl\"a, Finland}
\author{I.~Lazarus}
\affiliation{STFC Daresbury, Daresbury, Warrington WA4 4AD, United Kingdom}
\author{M.~Madurga}
\affiliation{Dept. of Physics and Astronomy, University of Tennessee, Knoxville, Tennessee 37996, US}
\author{N.~Marginean}
\affiliation{Horia Hulubei National Institute for R\&D in Physics and Nuclear Engineering, RO-077125 Bucharest, Romania}
\author{R.~Marginean}
\affiliation{Horia Hulubei National Institute for R\&D in Physics and Nuclear Engineering, RO-077125 Bucharest, Romania}
\author{C.~Mihai}
\affiliation{Horia Hulubei National Institute for R\&D in Physics and Nuclear Engineering, RO-077125 Bucharest, Romania}
\author{R.~E.~Mihai}
\affiliation{Horia Hulubei National Institute for R\&D in Physics and Nuclear Engineering, RO-077125 Bucharest, Romania}
\author{P.~Mosat}
\affiliation{Department of Nuclear Physics and Biophysics, Comenius University in Bratislava, 84248 Bratislava, Slovakia}
\author{J.~R.~Murias}
\affiliation{Grupo de F\`isica Nuclear, EMFTEL \& IPARCOS, 
Universidad Complutense de Madrid, 28040 Madrid, SPAIN }
\affiliation{Institut Laue-Langevin, CS 20156, 38042 Grenoble Cedex 9, France}
\author{E.~Nacher}
\affiliation{Instituto de F{\'i}sica Corpuscular, CSIC - Universidad de Valencia, E-46980, Valencia, Spain}
\author{A.~Negret}
\affiliation{Horia Hulubei National Institute for R\&D in Physics and Nuclear Engineering, RO-077125 Bucharest, Romania}
\author{R.~D.~Page}
\affiliation{Department of Physics, Oliver Lodge Laboratory, University of Liverpool, Liverpool L69 7ZE, United Kingdom}
\author{A.~Perea}
\affiliation{Instituto de Estructura de la Materia, CSIC, Serrano 113 bis, E-28006 Madrid, Spain}
\author{V.~Pucknell}
\affiliation{STFC Daresbury, Daresbury, Warrington WA4 4AD, United Kingdom}
\author{P.~Rahkila}
\affiliation{University of Jyv\"askyl\"a, Department of Physics, Accelerator laboratory, P.O. Box 35(YFL) FI-40014 University of Jyv\"askyl\"a, Finland}
\author{K.~Rezynkina}
\affiliation{KU Leuven, Instituut voor Kern- en Stralingsfysica, B-3001 Leuven, Belgium}
\affiliation{Universit{\'e} de Strasbourg, CNRS, IPHC UMR7178, F-67000, Strasbourg, France}
\author{V.~S{\'a}nchez-Tembleque}
\affiliation{Grupo de F\`isica Nuclear, EMFTEL \& IPARCOS, 
Universidad Complutense de Madrid, 28040 Madrid, SPAIN }
\author{K.~Schomacker}
\affiliation{Institut f{\"u}r Kernphysik, Universit{\"a}t zu K{\"o}ln, D-50937 K{\"o}ln, Germany}
\author{M.~Stryjczyk}
\affiliation{KU Leuven, Instituut voor Kern- en Stralingsfysica, B-3001 Leuven, Belgium}
\affiliation{University of Jyv\"askyl\"a, Department of Physics, Accelerator laboratory, P.O. Box 35(YFL) FI-40014 University of Jyv\"askyl\"a, Finland}
\author{C.~S{\"u}rder}
\affiliation{Institut f{\"u}r Kernphysik, Technische Universit{\"a}t Darmstadt, D-64289 Darmstadt, Germany}
\author{O.~Tengblad}
\affiliation{Instituto de Estructura de la Materia, CSIC, Serrano 113 bis, E-28006 Madrid, Spain}
\author{V.~Vedia}
\affiliation{Grupo de F\`isica Nuclear, EMFTEL \& IPARCOS, 
Universidad Complutense de Madrid, 28040 Madrid, SPAIN }
\author{N.~Warr}
\affiliation{Institut f{\"u}r Kernphysik, Universit{\"a}t zu K{\"o}ln, D-50937 K{\"o}ln, Germany}

\collaboration{IDS Collaboration}

\date{\today}

\begin{abstract}

Polonium isotopes having two protons above the shell closure at $Z=82$ show a wide variety of low-lying high-spin isomeric states across the whole chain. 
The structure of neutron-deficient isotopes up to $^{210}$Po ($N=126$) is well established as they are easily produced through various methods. However, there is not much information available for the neutron-rich counterparts for which only selective techniques can be used for their production. 
We report on the first fast-timing measurements of yrast states up to the 8$^+$ level in $^{214,216,218}$Po isotopes produced in the $\beta^-$ decay of $^{214,216,218}$Bi at ISOLDE, CERN. 
In particular, our new half-life value of 607(14)\,ps for the 8$_1^+$ state in $^{214}$Po is nearly 20 times shorter than the value available in literature and comparable with the newly measured half-lives of 409(16) and 628(25)\,ps for the corresponding 8$_1^+$ states in $^{216,218}$Po, respectively.
The measured $B(E2;8_1^+ \to 6_1^+)$ transition probability values follow an increasing trend relative to isotope mass, reaching a maximum for $^{216}$Po. The increase contradicts the previous claims of isomerism for the $8^+$ yrast states in neutron-rich $^{214}$Po and beyond. 
Together with the other measured yrast transitions, the $B(E2)$ values provide a crucial test of the different theoretical approaches describing the underlying configurations of the yrast band. 
The new experimental results are compared to shell-model calculations using the KHPE and H208 effective interactions and their pairing modified versions, showing an increase in configuration mixing when moving towards the heavier isotopes.

\end{abstract}

\maketitle

%%%%%%%%%%%%%%%%%%%%%% Introduction %%%%%%%%%%%%%%%%%%%%%%%%%%%%

The atomic nucleus, much like an atom itself, is composed of discrete particles, protons and neutrons, which fill energy levels according to quantum principles. In many ways, it behaves as a collection of individual nucleons in specific quantum states. 
 Nuclear isomers are excited meta-stable states with half-lives of the order of nanoseconds and longer. The isomerism is usually caused by large differences between initial and final states in physical properties such as total angular momentum, its projection on the symmetry axis ($K$-isomers), shape, seniority or the low energy difference between the states \cite{Gar2023,walker99}. 
By determining the decay energy, branching ratio of de-exciting transitions and half-life of the state in question, one can deduce the reduced transition probability (a measure of the decay strength usually expressed in Weisskopf units or W.u. \cite{bohr1998}) and compare it to the single-particle estimate (which assumes that the decay involves a single nucleon). The ratio of the experimental and single-particle values defines a hindrance (or enhancement) factor which is related to the underlying decay mechanism or the collective nature of a transition. 
The occurrence of isomeric states in the vicinity of doubly-magic shell closures represents one of the important benchmarks for testing the predictive power of the nuclear shell model using different residual nucleon-nucleon interactions \cite{Nai21}.  
 
In the region around the doubly-magic $^{208}_{82}$Pb$_{126}$, low-energy nuclear structure is often dominated by a relatively high-$j$ single-particle orbital (with $j\geq$ 7/2). Seniority ($\nu$) represents the number of particles that are not paired (i.e., they are not coupled to angular momentum $J=0$) and can be regarded as a good quantum number. Seniority isomers  arise due to the selection rules associated with the conservation of $\nu$ and are encountered in semi-magic nuclei because electric quadrupole ($E2$) transitions between $\nu$ = 2 states ($e.g.$ $8_1^+\to 6_1^+$) are small when the valence shell is close to half-filled \cite{Isacker_2011}. 

 Polonium isotopes, having two protons in the $h_{9/2}$ orbital above the $Z=82$ shell closure, are especially suitable for testing the seniority scheme across the long chain of isotopes, also crossing the $N=126$ shell closure. 
A large bulk of data exists in the literature on excited states and reduced transition probabilities $B(E2)$ in neutron-deficient $^{198-208}$Po isotopes. These nuclei can be studied through fusion-evaporation reactions with heavy ions \cite{Maj1990, Sto2023}, Coulomb excitation experiments \cite{Grahn_2016, Wrzosek_Lipska_2016} or $\beta^+/$EC decay \cite{bijnens1995}.
 These studies generally agree that the structure of the yrast states (defined as the states of lowest energy for a given angular momentum), including the $8^+$ seniority isomers, in the neutron-deficient polonium isotopes is dominated by the two-proton configuration $\pi(h_{9/2})^2$ coupled to the quadrupole vibrations of the underlying even-even $^{208}$Pb core \cite{Hau1976, neyens1997, Dracoulis_2016}. 

For isotopes heavier than $^{210}$Po, the high-$j$ neutron orbital $\nu(g_{9/2})$ starts to be filled. It is therefore expected to compete with the $\pi(h_{9/2})^2$ in determining the configuration of $0^+-8^+$ yrast states. The experimental studies of neutron-rich nuclei with $Z>82$ and $N>126$ are more challenging, as only very specific production techniques can be employed. One of the methods to populate excited states in such nuclei is through multi-nucleon transfer reactions such as $^{18}$O\,+\,$^{208}$Pb, used to study $^{210,212,214}$Po  \cite{astier2010,astier2011,kocheva2017,kocheva2017-2}. 

The previous nuclear structure study of $^{214}$Po \cite{astier2011} reported its yrast $8^+$ state half-life of $13(1)$\,ns and $B(E2;8_1^+ \to 6_1^+)=0.54(4)$\,W.u., underlining its isomeric character. 
Based on the resemblance between the excitation energies of yrast states in $^{214,216,218}$Po, the authors proposed that similar isomers might exist in $^{216,218}$Po, their main excitation mechanism being one-neutron-pair breaking. 

To reach isotopes beyond $^{214}$Po, high-energy fragmentation or spallation in direct or inverse kinematics are the only methods of choice  \cite{benzoni2012, morales2014}. In the past, by using proton-induced spallation in $^{232}$Th and $^{238}$U at ISOLDE, CERN, yrast states in $^{216,218}$Po up to $8^+$ have been populated through the $\beta^-$ decay of $^{216,218}$Bi \cite{kurpeta2000, dewitte2004}, however, half-life measurements of excited states were not performed. 

In this Letter, we report on the first fast-timing measurements of the yrast $2^+,4^+,6^+$ and $8^+$ states in $^{214,216,218}$Po populated in the $\beta^-$ decay of $^{214,216,218}$Bi at ISOLDE, CERN. The present results strongly disagree with the previous literature value of the $T_{1/2}(8_1^+)$ in $^{214}$Po \cite{astier2011}. The extracted $B(E2)$ values are compared to state-of-the-art shell-model calculations in order to understand the underlying structure of the yrast bands of $^{214,216,218}$Po leading to a comprehensive characterization of the single-particle excitations in the “northeast” region of $^{208}$Pb in the nuclear chart. 
We note that the detailed discussion of the experimental setup, decay schemes and newly-identified states in $^{214,216}$Po from the same experiment has been recently reported for the $\beta^-$ decay of $^{214}$Bi \cite{And21}  and $^{216}$Bi \cite{And24}.

%%%%%%%%%%%%%%%%%%%%%% Experimental results %%%%%%%%%%%%%%%%%%%%%%%%%%%%
\medskip
The $^{214,216,218}$Bi isotopes were produced at ISOLDE-CERN~\cite{ref:isolde_new} by bombarding a 50\,g/cm$^2$ thick UC$_x$ target with the 1.4~GeV proton beam delivered by the Proton Synchrotron Booster. 
The bismuth atoms released from the target were selectively ionized by the Resonance Ionization Laser Ion Source (RILIS) \cite{rilis}, accelerated to 50~keV and mass-separated using the ISOLDE High Resolution Separator.  
The bismuth ions were implanted on an aluminised Mylar$^\circledR$ tape at the center of the ISOLDE Decay Station (IDS) detection setup \cite{ids} equipped with a fast EJ232 plastic scintillator used as a $\beta$ detector, four HPGe Clover detectors for the detection of $\gamma$ rays in the daughter nuclei, and two small-volume (30.5\,cm$^3$) conic LaBr$_3$(Ce) detectors. This detection setup enabled the measurement of nuclear half-lives using the $\beta\gamma(t)$, $\gamma\gamma(t)$ and $\beta\gamma\gamma(t)$ fast-timing method \cite{mach2, mach1}, a well established technique at IDS employing a similar setup as described in Refs.~\cite{fraile, lica2016, lica2017-149Ba, lica2017-150Ba}. It covers half-lives between 10\,ps and 100\,ns by using fast-timing detectors such as LaBr$_3$(Ce) and plastic scintillators in coincidence with HPGe detectors. 
More details about the calibration procedure of the detectors is provided in the Supplemental Material \cite{supplemental}.

\begin{figure}[h!]
\centering
\includegraphics[width=0.49\textwidth]{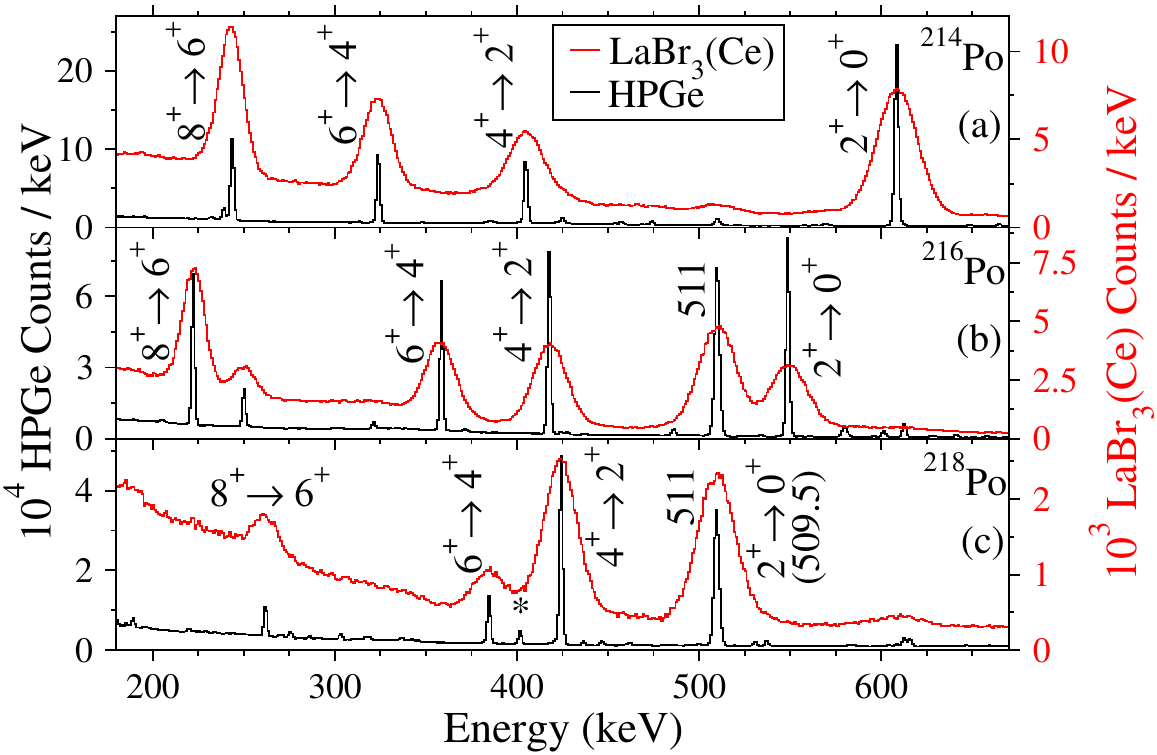}
\caption{\label{fig:spectra} $\beta$-gated $\gamma$-ray spectra recorded by the HPGe (black) and LaBr$_3$(Ce) (red) detectors following the $\beta^-$ decay of $^{214}$Bi (a), $^{216}$Bi (b) and $^{218}$Bi (c). The yrast transitions in $^{214,216,218}$Po are labeled. $\gamma$ rays from the $\beta$-decays of $^{84-88}$Br were also identified, originating from Ba$^{84-88}$Br$^+$ radioactive molecules produced in the target and matching mass 214, 216 or 218. The most intense contaminant visible in the region of interest is marked with the '*' symbol in panel (c) and originates from the $\beta$-decay of $^{87}$Kr.
}\end{figure}

Energy spectra recorded by the HPGe and LaBr$_3$(Ce) detectors in coincidence with $\beta^-$ events are shown in Fig.\,~\ref{fig:spectra}.
The production yields of $^{214,216,218}$Bi ions were measured to be $>2\times10^4$ (limited in order to reduce the count rate in the detectors), $1.5\times10^3$ and $2\times10^2$~ions/$\mu$C, considering an average proton current of 1.2, 1.6 and 1.8~$\mu$A and a total running time of 3, 6.5 and 13~h respectively.

\begin{figure}[h!]
\centering
\includegraphics[width=0.47\textwidth]{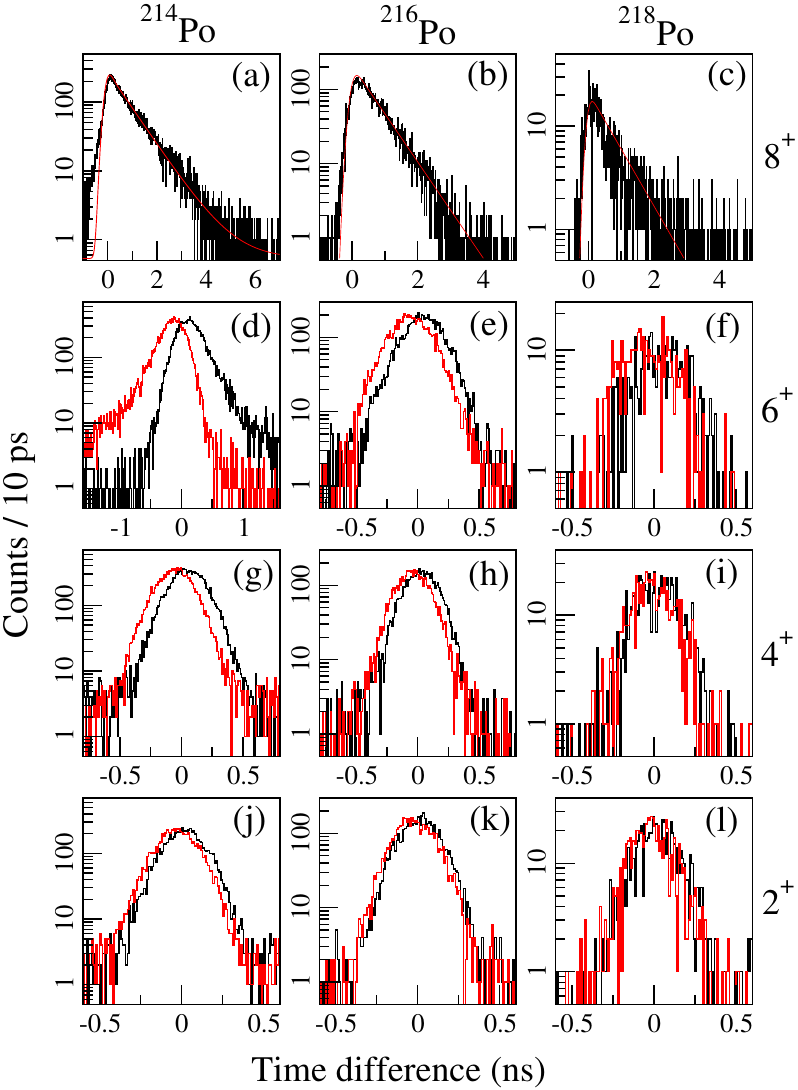}
\caption{\label{fig:8plus} (a-c) Background-subtracted time difference distributions  between $\beta$ particles recorded in the fast plastic scintillator and $\gamma$ rays directly depopulating the yrast $8^+$ states in $^{214,216,218}$Po, extracted from $\beta$-$\gamma_{La}$-$\gamma_{Ge}$ events.  (d-l) Delayed (black) and anti-delayed (red) background-subtracted time distributions between $\gamma$ rays directly feeding and de-exciting the yrast 2$^+$, 4$^+$ and 6$^+$ states in $^{214,216,218}$Po, extracted from $\gamma_{La}$-$\gamma _{La}$ events. The half-lives, reported in Table\,\ref{tab:be2}, were measured using the decay slope method (a-c) and the centroid-shift method (d-l) \cite{mach2, mach1}. The longer lived component (tail) in the $^{214}$Po 6$^+$ spectrum (d) originates from the 8$^+$ half-life due to an overlap in the LaBr3(Ce) energy spectra between the feeding (240 keV) and de-exciting (245 keV) transitions. However, it was excluded when evaluating the centroid shift of the distribution. }
\end{figure}

The half-lives of the $8_1^+$ states in $^{214,216,218}$Po were extracted through the $\beta\gamma\gamma(t)$ method by gating on $\beta$ particles in the plastic scintillator as the START signal and the $8_1^+ \rightarrow 6_1^+$ $\gamma$-rays in the LaBr$_3$(Ce) detectors as the STOP signal for the Time to Amplitude Converters. An extra gating condition was required on $\gamma$ rays in the HPGe detectors originating from any of the three transitions below the $6_1^+$ levels ($6_1^+ \to 4_1^+, 4_1^+ \to 2_1^+$ and $2_1^+ \to 0_1^+$) in order to further reduce the background of resulting time-difference distributions, shown in Fig.~\ref{fig:8plus} (a-c). 

The possible delayed contributions from high-lying long-lived levels feeding the $8_1^+$ states were investigated by HPGe gating on the $\gamma$-ray transitions from above. In the case of $^{214,216}$Po, we report half-lives of 73(7) and 155(14)\,ps for the $8_2^+$ 1824.5- \cite{And21} and 1802.6-keV \cite{kurpeta2000} levels, respectively. For $^{218}$Po, no long-lived states feeding the $8_1^+$ level were identified, either because of the lack of statistics or their absence. 
The  LaBr$_3$(Ce) detectors cannot resolve the transitions feeding (240 keV) and de-exciting (244 keV) the yrast 8$^+$ state in $^{214}$Po, see Fig.\,~\ref{fig:spectra} (a), however, the two contributions were taken into account for the half-life analysis.
The values of the half-lives reported in Table~\ref{tab:be2} are corrected for the contributions mentioned above and result from the convolution fit between a Gaussian (prompt response of the fast-timing detectors), an exponential decay and a flat background.

The time distributions for the decays of $2^+_1$, $4^+_1$, $6^+_1$ states in $^{214,216,218}$Po are shown in Fig.~\ref{fig:8plus} (d-l). The deduced half-lives are reported in Table~\ref{tab:be2} and have been measured through the $\gamma\gamma(t)$ method by extracting the timing information from the LaBr$_3$(Ce) detectors for $\gamma$ rays directly populating and de-populating each level. 

In the case of $^{214}$Po, a significant discrepancy is observed between the newly measured half-life value of $T_{1/2}(8_1^+)=607(14)$\,ps, and the previous one of 13(1)\,ns \cite{astier2011}. The latter was measured using the slope fitting of time-difference distributions from HPGe detectors, known for having a much slower and less accurate time response \cite{Cre10} compared to the LaBr$_3$(Ce) detectors \cite{fraile}.

%%%%%%%%%%%%%%%%%%%%%% Theory, discussion %%%%%%%%%%%%%%%%%%%%%%%%%%%%

\medskip
To link the newly measured data to the underlying nuclear structure, two effective interactions were used in the present work for the calculation of low-energy levels and $E2$ transition rates: 
(i) the recently-developed H208 effective interaction \cite{Nai19, Nai21} already used in the interpretation of $^{212,214,216}$Po new experimental data \cite{And21,Fer21,Kar22,And24};
(ii) the well-established Kuo-Herling interaction as modified by Warburton and Brown~\cite{War91}, denoted as KHPE. 
More details regarding the calculations are provided in the Supplemental Material~\cite{supplemental}. 

Additional modifications of the two interactions, denoted by H208-m and KHPE-m, were implemented within this work in order to reconcile calculated and experimental $B(E2)$ values but maintain the good agreement for other observables using the original interactions. Specifically, the diagonal and off-diagonal $1h_{9/2}$ proton-pairing matrix elements were reduced by 100\,keV in H208-m. In the case of the KHPE-m interaction, only a larger reduction of 200\,keV produced a similar improvement for $^{210-216}$Po. However, in $^{218}$Po, the $8_1^+$ state excitation energy becomes lower than the $6_1^+$ one and therefore the reduction was limited to 100\,keV in the case of $^{218}$Po. The $1h_{9/2} 0i_{13/2}$ matrix elements were not modified, as they had an insignificant effect on the studied $E2$ transitions.

\begin{figure}[h!]
\includegraphics[width=0.48\textwidth]{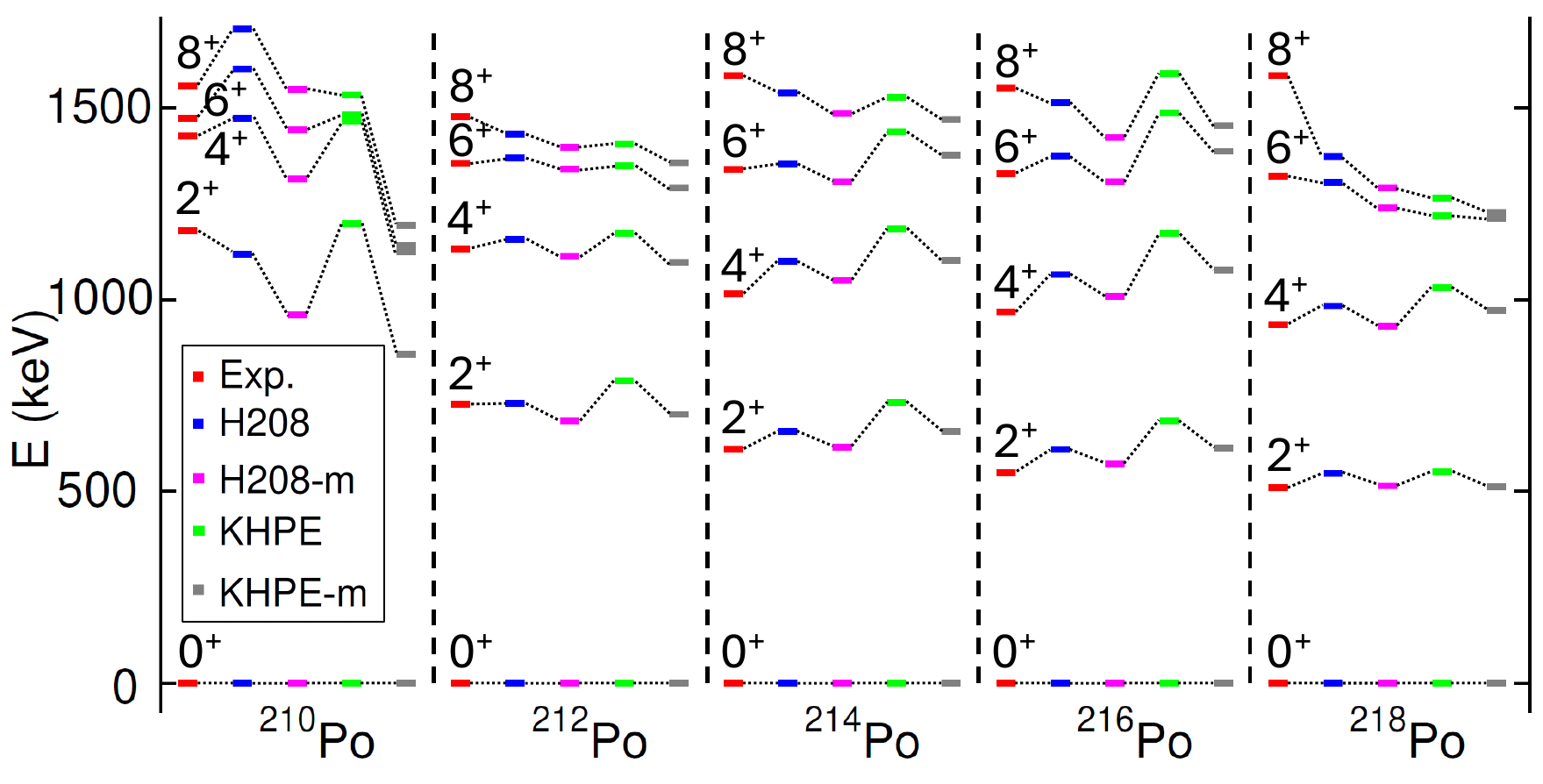}
\caption{The calculated $0^+-8^+$ yrast energy levels of even-even $^{210-218}$Po isotopes using H208, H208-m, KHPE and KHPE-m interactions, compared to the available experimental data \cite{nds210,nds212,nds214,nds216,nds218,astier2011}.}
\label{fig:energySM}
\end{figure}

%%%%%%%%%%%%%%%%%%%%%% Energy levels 

The calculated and experimental excitation energies 
of the $0^+-8^+$ yrast levels in $^{210-218}$Po isotopes are shown in Fig.~\ref{fig:energySM}. The comparison indicates that  the H208 interaction is in better agreement with the adopted experimental data \cite{nds212,nds214,nds216,nds218} than KHPE, except for $^{210}$Po \cite{nds210}. A common shortcoming is a compression by about 200~keV observed for the $8_1^+$ state of $^{218}$Po, which we attribute to the adopted truncation for that nucleus.

In order to investigate in detail the wave function structure of the yrast states, Table\,II in the Supplemental Material \cite{supplemental} reports their main components (with $\ge$ 10\% contribution), where we note the following points:
(i) all the yrast $2^+-8^+$ states in $^{210}$Po ($N=126$) have nearly pure proton $1h_{9/2}$ contribution;
(ii) for $N>126$ the contribution from the pure $1h_{9/2}$ diminishes, and the composition is more fragmented. The neutron $1g_{9/2}$ state starts to play a strong role in all the yrast states, $e.g.$ nearly 54\% in $0^+-6^+$ states in $^{212}$Po, and with a further admixture of $1f_{7/2}$ proton for the $8_1^+$ state. An even more fragmented wave function can be seen for $^{214-218}$Po, with the added contribution from the $0i_{11/2}$ neutron orbital.

%%%%%%%%%%%%%%%%%%%%%%%%%%%%  BE2

The calculated $B(E2; J^\pi \to (J-2)^\pi)$ reduced transition probabilities interconnecting the yrast $0^+-8^+$ states are compared to our new results and to available data from literature \cite{nds210,nds212,astier2011,kocheva2017,Tre21} in Fig.~\ref{fig:be2}, Table\,\ref{tab:be2} here and Table\,I in the Supplemental Material \cite{supplemental}. The results of the H208 and KHPE calculations are close to each other, with the exception of $^{218}$Po, where different truncations are implemented. Overall, they are within error bars with the new measurements, with few exceptions mentioned below.

\begin{table}[h]
\begin{tabular}{|c|c|c|c|c|c|c|c|}
\hline
        &   &  & \multicolumn{5}{c|}{   $B(E2;J^\pi \to (J-2)^\pi)$ (W.u.)} \\
\hline
  Nucl. &   $J^\pi$ & $T_{1/2}$ (ps)& Exp. &  H208  &  H208   & KHPE & KHPE \\ 
        &           & Exp.    &      &        &  -m     &      & -m   \\
\hline
           & $2_1^+$   &  13(5)   & 7(3)     & 13.6 & 14.3 & 12.3 & 13.2 \\
           & $4_1^+$   &  35(5)   & 18(3)    & 16.9 & 18.5 & 13.3 & 16.5 \\
$^{214}$Po & $6_1^+$   &  118(5)  & 16(1)    & 11.4 & 14.7 & 5.9 & 12.6  \\
           & $8_1^+$ &  607(14) & 11.3(3)  & 1.2  & 9.2  & 0.1 & 6.6\\
           &         &  \textbf{13(1)\,ns }& \textbf{0.54(4)}\cite{astier2011}  &  &   &  &   \\
\hline
           & $2_1^+$   &  11(5)   &  13(6)    & 18.1 & 18.9 & 17.7 & 18.7 \\
           & $4_1^+$   &  21(5)   &  26(6)    & 25.2 & 27.1 & 22.1 & 25.2 \\
$^{216}$Po & $6_1^+$   &  31(5)   &  37(6)    & 18.8 & 25.0 &  9.8 & 23.2 \\
           & $8_1^+$ &  409(16) &  24(1)    & 16.2 & 17.8 &  3.2 & 15.5 \\
\hline
           & $2_1^+$   & $<$15    &  $>$13   & 19.2 & 20.1 & 14.8 & 15.3   \\
           & $4_1^+$   &  $<$15   &  $>$33   & 29.9 & 31.5 & 21.5 & 22.9   \\
$^{218}$Po & $6_1^+$   &  20(8)   &  40(16)  & 28.3 & 35.0 &  2.8 &  3.2   \\
           & $8_1^+$ &  628(25) &  7.8(3)  &  8.5 & 16.2 &  1.0 &  0.003 \\
\hline
\end{tabular}
\caption{\label{tab:be2} Experimental $T_{1/2}$ and $B(E2)$ values in $^{214-218}$Po measured in the present work and Ref.\,\cite{astier2011} (bold), compared to calculated $B(E2)$s using various effective interactions: H208, KHPE, and their pairing-modified versions. }
\end{table}

\begin{figure}[h]
\centering
\includegraphics[width=0.47\textwidth]{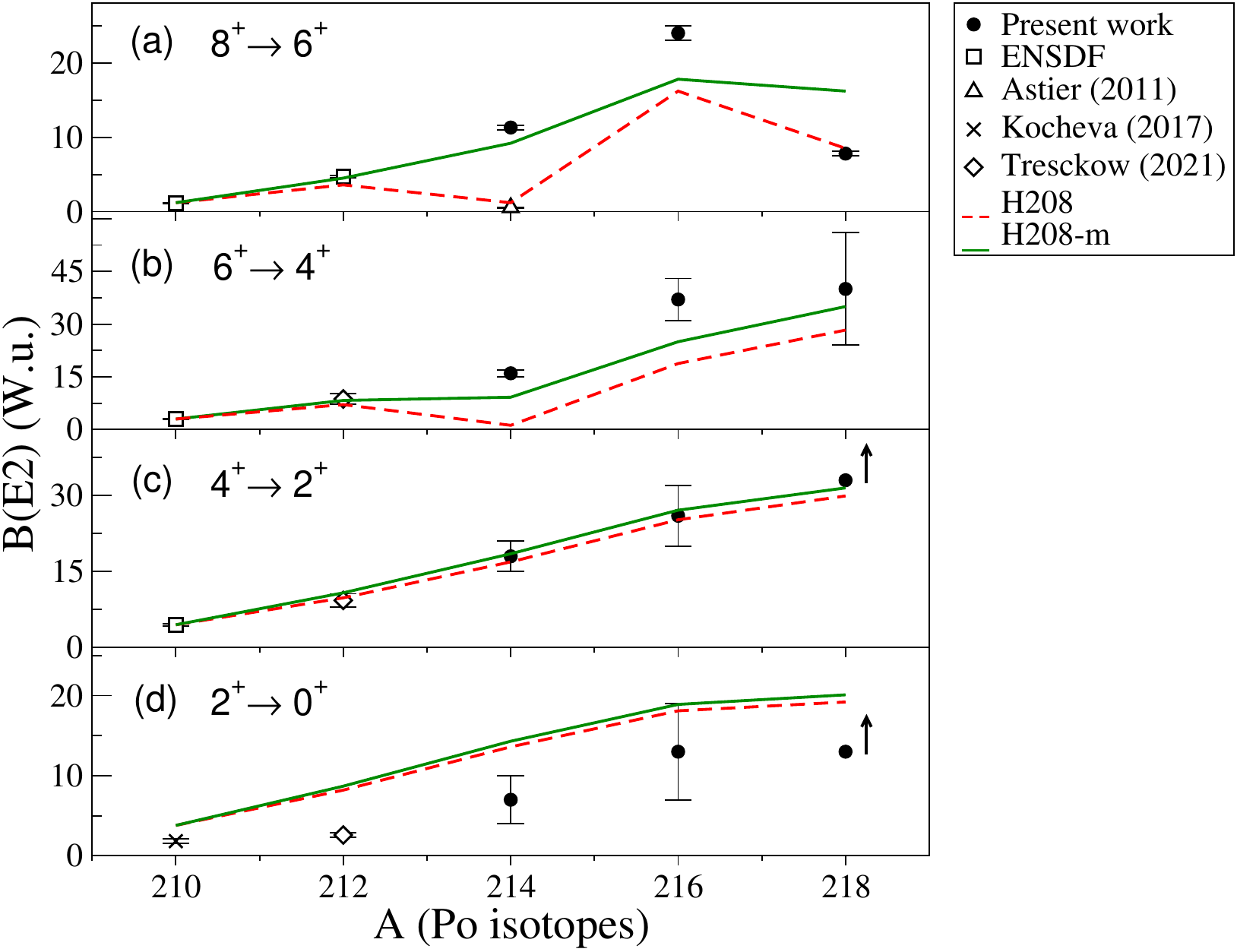}
\caption{\label{fig:be2} $B(E2; J^\pi \rightarrow (J-2)^\pi)$ values for transitions between $^{214,216,218}$Po yrast states measured in the present work (full circles with uncertainties) and from adopted experimental values \cite{nds210,nds212} or the most recent experimental results \cite{astier2011,kocheva2017,Tre21}. The comparison with shell-model calculations using the H208 interaction initial and pairing-modified versions (colored lines) is presented. Lower limits for the experimental $B(E2)$ values are indicated with an arrow pointing up. All the values are reported in Table\,\ref{tab:be2}. }
\end{figure}

We first note the over-estimation of the $B(E2,2_1^+ \to 0_1^+)$ values in $^{210,212,214}$Po isotopes by a factor of two. However, the calculated $B(E2;4_1^+\to 2_1^+)$ and $B(E2;6_1^+\to 4_1^+)$ values are very well reproduced by the shell-model calculations, shown in Table\,\ref{tab:be2}, which suggests that the correct structure of the $2_1^+$ state wave function was taken into account. 
This $B(E2,2_1^+ \to 0_1^+)$ discrepancy was already explained in Refs.\,\cite{War91, Kar19} by the absence of the contribution from particle-hole core excitations in the $0_1^+$ wave function and later, the $\alpha$-cluster model was proposed to address this issue \cite{Tre21}.

The most striking discrepancy is observed for the $B(E2;8_1^+ \to 6_1^+)$ values in $^{214}$Po, where our newly measured value is at least 10 times larger than the original shell-model calculated one, questioning the isomeric nature of the $8_1^+$ state in $^{214}$Po and excluding any hindrance originating from seniority selection rules. This represents the first experimental indication of an increased fragmentation of the yrast states wavefunctions. 
To understand the origin of this inconsistency between theory and our new measurements, the modified effective interactions were employed. As a consequence, there was a remarkable increase of the calculated $B(E2;8_1^+\to 6_1^+)$ value in $^{214}$Po, from 1.2 to 9.2\,W.u. in H208-m and from 0.1 to 6.6 W.u. in KHPE-m, close to the measured value of 11.3(3)\,W.u., as shown in Fig.\,\ref{fig:be2} and Table\,\ref{tab:be2}. 

\begin{figure}[h]
\centering
\includegraphics[width=0.50\textwidth]{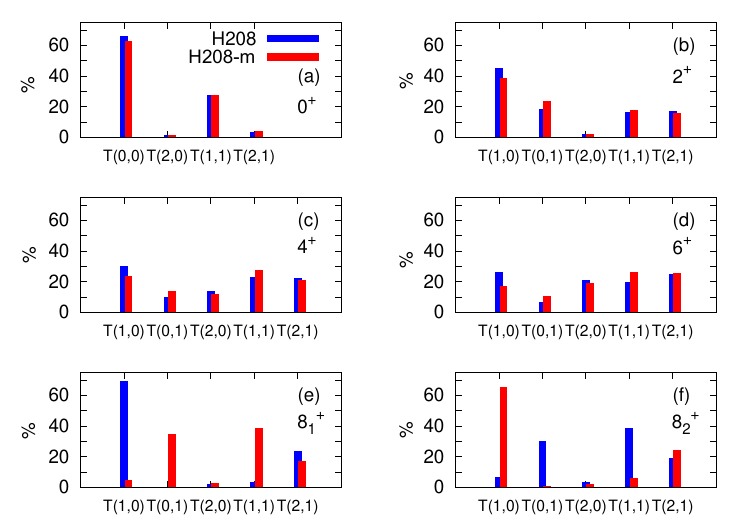}
%\scalebox{.7}{\includegraphics{wf.pdf}}
\caption{The wave-function components of $^{214}$Po states calculated in seniority scheme where $T(x,y)$ represents the number of neutron ($x$) or proton ($y$) pairs being broken. The initial and pairing-modified versions of the H208 effective interaction are employed within the \textsc{Nathan} code. }
\label{fig:seniority}
\end{figure}

To provide a complementary insight into the effect of the aforementioned pairing modifications on the wave functions structure of the lowest states in $^{214}$Po, the same calculations are performed in the seniority $(\nu)$ scheme using the $j$-coupled \textsc{Nathan} code \cite{Nathan}. The results obtained using H208 and H208-m are displayed in Fig.\,\ref{fig:seniority}, where $T(x,y)$ represents the number of neutron ($x$) or proton ($y$) pairs being broken ($e.g.$ the $0_1^+$ state wavefunction is dominated by a 66\% contribution from $T(0,0)$ which corresponds to $\nu=0$, and the $2_1^+$ by $45\%$ of $T(1,0)$ corresponding to $\nu=2$). No important change between the two versions can be observed in the seniority structure of the $0_1^+, 2_1^+, 4_1^+$ and $6_1^+$ states. 
The effect of reducing the $1h_{9/2}$ proton pairing matrix elements is apparent on the structure of the two lowest $8^+$ states. In H208, the main wave function component of the $8_1^+$ state was characterized by seniority $\nu=2$ (one neutron pair broken). The $8_2^+$ state was initially predicted very close in energy to the $8_1^+$ state, connected by a strong $E2$ transition to the $6_1^+$ state and characterized by a mixing of $\nu$=2, 4 and 6. Reducing the pairing in H208-m leads to an inversion of the $8_{1,2}^+$ states, the $\pi (1h_{9/2} 1f_{7/2})$ state comes below the $\nu g_{9/2}^4$ $8^+$ one, resulting in a good agreement between experiment and calculations. 
In this regard, it is important to notice that similar discrepancies between calculated and experimental $B(E2)$ values have been found before in $^{136}$Sn \cite{Sim14} and Pb isotopes \cite{Got12}. All these cases share a common point that the inclusion of the effective three-body forces via the adjustment of the two-body matrix elements improves the agreement with the experimental data \cite{Zuk03}.

%%%%%%%%%%%%%%%%%%%%%%%%%%%%  BE2 trends

The increasing trend of the experimental $B(E2; J^\pi \to (J-2)^\pi)$ values for transitions from the $J=2, 4, 6, 8$ states, shown in Fig.~\ref{fig:be2}, was predicted theoretically in neutron-rich Po isotopes up to $A=216$, with the exception of a dip in $^{214}$Po, for the $8_1^+ \to 6_1^+$ transition (see Figs.\,6-9 from Ref.\,\cite{Nai21}). 
The trend was explained through the increase of the collectivity and quadrupole correlations with respect to the neutron number, leading to strong electromagnetic strengths. 

The sudden decrease of the experimental $B(E2;8_1^+ \to 6_1^+)$ for $^{218}$Po is also reproduced theoretically, despite the adopted truncation in this nucleus. This can be attributed to the contribution from $i_{13/2}$ and $j_{15/2}$ proton and neutron orbitals in the structure of the $8_1^+$ and $6_1^+$ states. A strong transition probability of 27\,W.u. is calculated when excluding the aforementioned orbitals. Their inclusion determines a reduction of quadrupole correlations due to pairing, leading to the $B(E2;8_1^+ \to 6_1^+)$ decrease observed in $^{218}$Po.

%%%%%%%%%%%%%%%%%%%%%% Conclusions %%%%%%%%%%%%%%%%%%%%%%%%%%%%

\medskip
In conclusion, the $B(E2)$ transition probabilities of yrast states up to the $8^+$ level in $^{214,216,218}$Po isotopes populated in the $\beta^-$ decay of $^{214,216,218}$Bi are established for the first time through fast-timing measurements at the ISOLDE Decay Station. 
The comparison with shell-model calculations using the standard versions of the H208 and KHPE effective interactions shows reasonable agreement for the lower-lying transitions. 
A significant disagreement was observed in the case of the $8_1^+ \to 6_1^+$ transitions, especially for $^{214}$Po where a much higher transition probability was measured ($B(E2)= 11.3(3)$\,W.u. instead of 0.54(4)\,W.u. \cite{astier2011}), disproving its previously-claimed isomeric character. Similarly, based on the large $B(E2)$ values, the $8_1^+$ states in $^{216-218}$Po should not be considered isomeric in the context of the seniority scheme as there is no characteristic hindrance observed in their decays. This latter observation is indicative of an increased fragmentation of the yrast states wavefunctions. 
An improved agreement between theory and the present measurement was achieved after reducing the pairing strength of the interactions, effectively inverting the lowest predicted $8^+$ states and confirming the two-proton configuration $\pi (1h_{9/2} 1f_{7/2})$ of the yrast $8^+$ state dominated by quadrupole correlations. 
The new findings can serve as a guideline for future analysis, providing a clear indication of the nature of $8_1^+$ states in  neutron-rich polonium isotopes. Additionally, they serve as an extremely valuable input for the different theoretical approaches and constitute a stringent test for the effective interactions.

%%%%%%%%%%%%%%%%%%%%%% Acknowledgements %%%%%%%%%%%%%%%%%%%%%%%%%%%%

\medskip
We would like to thank B.A.~Brown, A.~Gargano and the late H.~Grawe for the inspiring discussions and the ISOLDE staff for the support provided during the experiment.   
We acknowledge support for this work by 
the Romanian IFA grant CERN/ISOLDE and Nucleu project No. PN 23 21 01 02,
the Slovak Research and Development Agency (Contract No. APVV-22-0282) and the Slovak grant agency VEGA (Contract No. 1/0651/21),
the Research Foundation Flanders (FWO, Belgium), BOF KU Leuven (C14/22/104), fWO and F.R.S.-FNRS under the Excellence of Science (EOS 40007501) programme, 
the Spanish MCIN/AEI/10.13039/ 501100011033 under grants RTI2018-098868-B-I00 and PID2021-126998OB-I00, 
the Academy of Finland project No. 354968,
the German BMBF under contract 05P21PKCI1 and Verbundprojekt 05P2021,
and the United Kingdom Science and Technology Facilities Council (STFC) through the grants ST/P004598/1 and ST/V001027/1.

\bibliography{214-218Po}

\end{document}

\title{Supplemental Material: On the nature of yrast states in neutron-rich polonium isotopes}
\maketitle
\subsection{Detector calibrations and corrections}
The absolute $\beta$ and $\gamma$-ray efficiencies for the detection setup of the present experiment were: ~15\% for $\beta$ electrons recorded in the fast plastic scintillator, 6\% for $\gamma$ rays at 300\,keV and 3\% at 500 keV for the two LaBr$_3$(Ce) detectors, 9\% at 300\,keV and 7\% at 500\,keV for the four HPGe detectors. 
The digital processing of the energy signals provides resolutions at 1.3~MeV of the order of 2.3~keV for the HPGe and 40~keV for the LaBr$_3$(Ce) detectors. 
The HPGe detectors' efficiency when gating on all three transitions below the $6_1^+$ level ($6_1^+ \to 4_1^+, 4_1^+ \to 2_1^+$ and $2_1^+ \to 0_1^+$) is $\sim$23\%, corresponding to the sum of the three individual absolute efficiency values for each $\gamma$-ray energy.

The first step of the fast-timing analysis involved the walk correction for the $\beta$ detector. 
Although the $\beta$ detector is thin and has a time response relatively independent of the incident $\beta$ energy, there is still a residual time dependence that needs to be accounted for. The correction was performed relative to an intense prompt peak in the LaBr$_3$(Ce) spectra for both calibration and experiment data sets. 
Afterwards, the curves representing the centroid positions of time spectrum vs. $\gamma$-ray energy deposited in the the LaBr$_3$(Ce) crystals were extracted for each full energy peak after subtracting the background close to it.  
The timing information was corrected on an event-by-event basis after each iteration using the polynomial fit of the energy-vs-time walk curve determined before by using the timing information of transitions in $^{138}$Ba which de-excite short-lived levels that are directly fed in the $\beta$ decay of implanted $^{138}$Cs \cite{138Ba, 138Ba-timing}. The final time-energy walk curves were extracted using HPGe gated coincidences between the plastic scintillator and the LaBr$_3$(Ce) detectors in the triple $\beta\gamma\gamma(t)$-event data set, showing an overall variation of $\pm5$\,ps compared to a flat response. 

\subsection{Shell-model calculations}
The two different shell-model calculations were performed including the same model space spanning the $0h_{9/2}$, $1f_{7/2}$, $1f_{5/2}$, $2p_{3/2}$, $2p_{1/2}$, $0i_{13/2}$ orbitals for protons, and the $1g_{9/2}$, $0i_{11/2}$, $1g_{7/2}$, $2d_{5/2}$, $2d_{3/2}$, $3s_{1/2}$, $0j_{15/2}$ orbitals for neutrons, above the $^{208}$Pb doubly magic core. 
The proton and neutron single-particle energies were extracted from experimental studies on $^{209}$Bi and $^{209}$Pb \cite{nds209}, respectively. The following diagonalization and truncation schemes were applied:

(a) The H208 diagonalization of the matrices was achieved using the \textsc{Antoine} shell-model code \cite{Cau99,Cau05}, where the full configuration space was included for the $^{210-216}$Po isotopes. A truncation scheme was adopted in $^{218}$Po, by including the excitations up to $6p-6h$ excitations, with some limitations in the occupation of the $0j_{15/2}$ neutron and the $0i_{13/2}$ proton orbitals. Despite this truncation, the diagonalization of the $^{218}$Po matrix, with its size of the order of $5 \times 10^9$ is very challenging. 

(b) The KHPE diagonalizations were performed using the \textsc{Kshell} code~\cite{Shi19}. Similarly to H208, calculations up to $^{216}$Po were done in the full space, while the calculation of $^{218}$Po needed truncation. This truncation was slightly different from the one used for H208. First, the two valence protons were limited only to the $0h_{9/2}$, $1f_{7/2}$ and $0i_{13/2}$ orbitals. Secondly, the excitations of the 8 valence neutrons were limited to maximum of four in the $0i_{11/2}$ and maximum of two neutrons in the $2d_{5/2}, 2d_{3/2}, 3s_{1/2}, 0j_{15/2}$ orbitals. Finally, an additional limitation on the joint occupation of the high-$j$ orbitals $\pi 0i_{13/2}$ and $\nu 0j_{15/2}$ was necessary to limit the calculation dimensions to $2.6 \times 10^9$. 

To facilitate the comparison between the two shell-model calculations, the same effective charges are used, $e_n=0.85e$ for neutrons and $e_p=1.5e$ for protons. 

To complement the presently reported experimental $B(E2)$ values and further validate the modifications implemented to the effective interactions, Table\,\ref{tab:be2SM} includes experimental $B(E2)$ values for transitions between yrast states in $^{210,212}$Po from literature \cite{nds210,nds212,kocheva2017,Tre21} compared to calculated $B(E2)$s. 

The principal wave function components ($\ge 10\%$) calculated using the H208 and H208-m effective interactions for yrast states in $^{210,212,214,216,218}$Po isotopes are listed in Table\,\ref{tab:wavefunctions}.

\begin{table*}[t]
\begin{tabular}{|c|c|c|c|c|c|c|}
\hline
        &   & \multicolumn{5}{c|}{   $B(E2;J\to J-2)$ (W.u.)} \\
\hline
  Nucleus &   $J^\pi$ &  Exp. &  H208  &  H208-m   & KHPE & KHPE-m \\ 
            &         &      &   &    &  &   \\
\hline
            & $2_1^+$ &   1.83(28) \cite{kocheva2017}, 0.56(12) \cite{nds210}  & 3.8  & 3.8 &3.5&3.5\\
            & $4_1^+$ &   4.46(18) \cite{nds210}  & 4.5 & 4.5 & 4.5&4.5\\
$^{210}$Po  & $6_1^+$ &   3.05(9) \cite{nds210}  & 3.0   &  3.0 &3.0&3.0\\
            & $8_1^+$ & 1.12(4) \cite{nds210}  & 1.2   & 1.2 &  1.2 &1.2\\
\hline
            & $2_1^+$ &  2.57$(^{+0.38}_{-0.29})$ \cite{nds212}, 2.6(3) \cite{Tre21} & 8.2   & 8.7  &6.6&7.8\\
            & $4_1^+$ &  9.3(13) \cite{Tre21} &  9.8 & 10.8  &7.9&9.8\\
$^{212}$Po  & $6_1^+$ &  13.2$(^{+4.9}_{-2.9})$ \cite{nds212}, 8.7(15) \cite{Tre21} &  7.1   & 8.3 &5.5&7.5\\
            & $8_1^+$ & 4.56(12) \cite{nds212}, 4.6(1) \cite{Tre21} &  3.6 & 4.5  &2.5&3.9 \\
\hline
\end{tabular}
\caption{\label{tab:be2SM} Experimental $B(E2)$ values for transitions between yrast states in $^{210,212}$Po  from literature \cite{nds210,nds212,kocheva2017,Tre21}, compared to calculated $B(E2)$s using the effective interactions employed in the present work: H208, KHPE, and their pairing-modified versions.}
\end{table*}

\begin{table*}[t]
\begin{center}
\begin{tabular}{|c|c|c|c|}
\hline
  Nucleus & $J^\pi$ &  H208 & H208-m \\
\hline
          & $0_1^+$&   76\% $\pi 1h_{9/2}^2$ + 14\% $\pi 1i_{13/2}^2$& 77\% $\pi 1h_{9/2}^2$ + 15\% $\pi 1i_{13/2}^2$\\
          & $2_1^+$ &     95\% $\pi 1h_{9/2}^2$& 95\% $\pi 1h_{9/2}^2$\\
           & $4_1^+$ &   98\% $\pi 1h_{9/2}^2$ & 98\% $\pi 1h_{9/2}^2$\\
$^{210}$Po  & $6_1^+$ &     99\% $\pi 1h_{9/2}^2$& 99\% $\pi 1h_{9/2}^2$\\
            & $8_1^+$ &    99\% $\pi 1h_{9/2}^2$& 99\% $\pi 1h_{9/2}^2$\\
\hline
          & $0_1^+$ &  46\% $\pi 1h_{9/2}^2\otimes \nu 1g_{9/2}^2$& 46\% $\pi 1h_{9/2}^2\otimes \nu 1g_{9/2}^2$\\
          & $2_1^+$ &  53\% $\pi 1h_{9/2}^2\otimes \nu 1g_{9/2}^2$&53\% $\pi 1h_{9/2}^2\otimes \nu 1g_{9/2}^2$\\
           & $4_1^+$ &  57\% $\pi 1h_{9/2}^2\otimes \nu 1g_{9/2}^2$&56\% $\pi 1h_{9/2}^2\otimes \nu 1g_{9/2}^2$\\
$^{212}$Po  & $6_1^+$ &  58\% $\pi 1h_{9/2}^2\otimes \nu 1g_{9/2}^2$&56\% $\pi 1h_{9/2}^2\otimes \nu 1g_{9/2}^2$\\
            & $8_1^+$ & 54\% $\pi 1h_{9/2}^2\otimes \nu 1g_{9/2}^2$+ 11\%  $\pi 1f_{7/2}^2\otimes \nu 1g_{9/2}^2$& 47\% $\pi 1h_{9/2}^2\otimes \nu 1g_{9/2}^2$ + 15\%  $\pi 1h_{9/2}1f_{7/2}\otimes \nu 1g_{9/2}^2$\\
\hline
        & $0_1^+$ &  22\% $\pi 1h_{9/2}^2\otimes \nu 1g_{9/2}^4$ + 12\% $\pi 1f_{7/2}^2\otimes \nu 1g_{9/2}^4$&
                21\% $\pi 1h_{9/2}^2\otimes \nu 1g_{9/2}^4$ + 12\% $\pi 1f_{7/2}^2\otimes \nu 1g_{9/2}^4$ \\
          & $2_1^+$ & 20\% $\pi 1h_{9/2}^2\otimes \nu 1g_{9/2}^4$+ 11\% $\pi 1f_{7/2}^2\otimes \nu 1g_{9/2}^4$& 
          19\% $\pi 1h_{9/2}^2\otimes \nu 1g_{9/2}^4$+ 10\% $\pi 1f_{7/2}^2\otimes \nu 1g_{9/2}^4$
          \\
    & $4_1^+$ &  20\% $\pi 1h_{9/2}^2\otimes \nu 1g_{9/2}^4$+ 12\% $\pi 1f_{7/2}^2\otimes \nu 1g_{9/2}^4$         &          18\% $\pi 1h_{9/2}^2\otimes \nu 1g_{9/2}^4$+ 11\% $\pi 1f_{7/2}^2\otimes \nu 1g_{9/2}^4$\\
$^{214}$Po & $6_1^+$ & 23\% $\pi 1h_{9/2}^2\otimes \nu 1g_{9/2}^4$+ 14\% $\pi 1f_{7/2}^2\otimes \nu 1g_{9/2}^4$ &   18\% $\pi 1h_{9/2}^2\otimes \nu 1g_{9/2}^4$+ 13\% $\pi 1f_{7/2}^2\otimes \nu 1g_{9/2}^4$  \\
       & $8_1^+$ &  34\% $\pi 1h_{9/2}^2\otimes \nu 1g_{9/2}^4$+ 16\% $\pi 1f_{7/2}^2\otimes \nu 1g_{9/2}^4$&   28\% $\pi 1h_{9/2}1f_{7/2}\otimes \nu 1g_{9/2}^4$+ 11\% $\pi 1h_{9/2}1f_{7/2}\otimes \nu 1g_{9/2}^2 1i_{11/2}^2$ \\
   & $8_2^+$   & 26\% $\pi 1h_{9/2}1f_{7/2}\otimes \nu 1g_{9/2}^4$+ 10\% $\pi 1f_{7/2}^2\otimes \nu 1g_{9/2}^2 1i_{11/2}^2$&  33\% $\pi 1h_{9/2}^2\otimes \nu 1g_{9/2}^4$+ 16\% $\pi 1f_{7/2}^2\otimes\nu 1g_{9/2}^4$ \\
\hline
            & $0_1^+$ &  13\% $\pi 1h_{9/2}^2\otimes \nu 1g_{9/2}^4 1i_{11/2}^2$ &  12\% $\pi 1h_{9/2}^2\otimes \nu 1g_{9/2}^4 1i_{11/2}^2$ \\
           & $2_1^+$ &    13\% $\pi 1h_{9/2}^2\otimes \nu 1g_{9/2}^4 1i_{11/2}^2$ &12\% $\pi 1h_{9/2}^2\otimes \nu 1g_{9/2}^4 1i_{11/2}^2$ \\
           & $4_1^+$ &     12\% $\pi 1h_{9/2}^2\otimes \nu 1g_{9/2}^4 1i_{11/2}^2$ &11\% $\pi 1h_{9/2}^2\otimes \nu 1g_{9/2}^4 1i_{11/2}^2$ \\
$^{216}$Po & $6_1^+$ &    10\% $\pi 1h_{9/2}^2\otimes \nu 1g_{9/2}^4 1i_{11/2}^2$ &10\% $\pi 1h_{9/2}^2\otimes \nu 1g_{9/2}^4 1i_{11/2}^2$ \\
            & $8_1^+$ & 12\% $\pi 1h_{9/2}1f_{7/2}\otimes \nu 1g_{9/2}^4 1i_{11/2}^2$ &16\% $\pi 1h_{9/2}1f_{7/2}\otimes \nu 1g_{9/2}^4 1i_{11/2}^2$\\
\hline
	 & $0_1^+$ & 13\% $\pi 1h_{9/2}^2\otimes \nu 1g_{9/2}^6 1i_{11/2}^2$+ 11\% $\pi 1f_{7/2}^2\otimes \nu 1g_{9/2}^6 1i_{11/2}^2$&
	 12\% $\pi 1h_{9/2}^2\otimes \nu 1g_{9/2}^6 1i_{11/2}^2$+ 11\% $\pi 1f_{7/2}^2\otimes \nu 1g_{9/2}^6 1i_{11/2}^2$
	 \\
           & $2_1^+$ & 13\% $\pi 1h_{9/2}^2\otimes \nu 1g_{9/2}^6 1i_{11/2}^2$+ 11\% $\pi 1f_{7/2}^2\otimes \nu 1g_{9/2}^6 1i_{11/2}^2$  & 
           12\% $\pi 1h_{9/2}^2\otimes \nu 1g_{9/2}^6 1i_{11/2}^2$+ 11\% $\pi 1f_{7/2}^2\otimes \nu 1g_{9/2}^6 1i_{11/2}^2$
           \\
           & $4_1^+$ &   12\% $\pi 1h_{9/2}^2\otimes \nu 1g_{9/2}^6 1i_{11/2}^2$+ 12\% $\pi 1f_{7/2}^2\otimes \nu 1g_{9/2}^6 1i_{11/2}^2$ & 
           11\% $\pi 1h_{9/2}^2\otimes \nu 1g_{9/2}^6 1i_{11/2}^2$+ 12\% $\pi 1f_{7/2}^2\otimes \nu 1g_{9/2}^6 1i_{11/2}^2$
           \\
$^{218}$Po & $6_1^+$ &    12\% $\pi 1h_{9/2}^2\otimes \nu 1g_{9/2}^6 1i_{11/2}^2$+ 11\% $\pi 1f_{7/2}^2\otimes \nu 1g_{9/2}^6 1i_{11/2}^2$ & 
  10\% $\pi 1f_{7/2}^2\otimes \nu 1g_{9/2}^61i_{11/2}^2$

\\
            & $8_1^+$ & 12\% $\pi 1h_{9/2}1f_{7/2}\otimes \nu 1g_{9/2}^6 1i_{11/2}^2$  & 
            19\% $\pi 1h_{9/2}1f_{7/2}\otimes \nu 1g_{9/2}^6 1i_{11/2}^2$ 
            \\
\hline
\end{tabular}
\end{center}
\caption{\label{tab:wavefunctions} The principal wave function components ($\ge 10\%$) for yrast states in $^{210,212,214,216,218}$Po isotopes, calculated using the two versions of the effective interaction: H208 (initial) and H208-m (reduced pairing).}
\end{table*}

\bibliography{214-218Po}